\documentclass[epsfig,onecolumn,12pt,a4paper]{article}
\usepackage{graphicx}
\usepackage{amsmath}

\input epsf.sty
\input{tcilatex}
\begin{document}

\title{An exact solution to the Extended Hubbard Model in $2D$ for finite
size system}
\author{S. Harir$^{1\thanks{%
harirsaid@gmail.com}},$ M. Bennai$^{1,3\thanks{%
bennai\_idrissi@yahoo.fr}}$ and Y. Boughaleb$^{1,2,4\thanks{%
yboughaleb@yahoo.fr}}$ \\
$^{1}${\small \ Laboratoire de Physique de la Mati\`{e}re Condens\'{e}e,}\\
{\small Facult\'{e} des Sciences Ben M'Sik, Universit\'{e} Hassan
II-Mohammedia Casablanca,\ Morocco. }\\
$^{2}${\small \ LPMC, Facult\'{e} des Sciences d' El Jadida, Universit\'{e}
Chouaib Doukkali, Morocco.}\\
$^{3}${\small \ Groupement National de Physique des Hautes Energies,
LabUFR-PHE, Rabat, Morocco.}\\
$^{4}$ {\small Hassan II Academy of Sciences and Technology, Morocco}}
\date{}
\maketitle

\begin{abstract}
An exact analytical diagonalization is used to solve the two dimensional
Extended Hubbard Model for system with finite size. We have considered an
Extended Hubbard Model (EHM) including on-site and off-site interactions
with interaction energy $U$ and $V$ respectively, for square lattice
containing $4\times 4$ sites at one-eighth filling with periodic boundary
conditions, recently treated by Kovacs et al \cite{kovacs}. Taking into
account the symmetry properties of this square lattice and using a
translation linear operator, we have constructed a $r-space$ basis, only,
with $85$ state-vectors which describe all possible distributions for four
electrons in the $4\times 4$ square lattice. The diagonalization of the $%
85\times 85$ matrix energy allows us to study the local properties of the
above system as function of the on-site and off-site interactions energies,
where, we have shown that the off-site interaction encourages the existence
of the double occupancies at the first exited state and induces
supplementary conductivity of the system.


\textbf{Pacs numbers:} 71.10, -W.75.10.Jm, 72.15.Nj
\end{abstract}

\newpage

\section{Introduction}

In the last years, the Extended Hubbard Model (EHM) was introduced to
explain some interesting phenomena including metal-insulator transition \cite%
{mott1,mott2,mott3,mott4,mott5}, antiferromagnetism \cite%
{Sanna,Nie,Antif3,antij}, and high-$T_{c}$ superconductivity \cite%
{Mc,Supra2,Halb}. This EHM is a standard simple model of interacting
itinerant electrons in a solid \cite{lieb,and,zhn,tak,zim,aaa}. Although
this model is too idealized to be regarded as a quantitatively reliable
model of real solids, it contains physically essential features of
interacting itinerant electron systems. Despite being one of the most
studied models to describe strongly correlated electrons system, many
questions concerning the Hubbard model remain as open problems. In \cite%
{applied}, we have applied the Self Consistent Random Phase Approximation
(SCRPA) \cite{jemai,RPA} to solve the EHM in $1D$, where, we have shown that
this approach treat the correlations of closed chains with a rigours manner.
The behaviour of our SCRPA ground state and gap energies shows that the
repulsive off-site interaction between the electrons of the neighbouring
atoms induces supplementary conductivity, since, the SCRPA energy gap
vanishes when the closed chains of the EHM are governed by a strong
repulsive on-site and an intermediate repulsive off-site interactions. But,
due to the restricted motion along one direction in space, the Hubbard model
chain doesn't exhibit any ferromagnetic feature. Thus, the $1D$ Hubbard is a
nice prototype to describe, only, the $1D$ classical conductors.

Stimulated by the discovery of high-$T_{c}$ superconductivity in the cuprate
plans, the $2D$ Hubbard model have attracted a great attention in the recent
two decades \cite{ben,ste}. In spite of its simple description in square
lattice, it is not obvious to solve the 2D Hubbard model in the general
case. The exact solution of $2D$ Hubbard model is not still reached, but, a
great variety of approximate treatments have been proposed \cite{kot,het}.

Very recently, E. Kovacs et al. \cite{kovacs} proposed an exact solution of
the Usual Hubbard Model for $4sites\times 4sites$ cluster with low
concentration, where each cluster presents four electrons. The dynamics of
these electrons is described by the usual kinetic of electrons with hopping
energy $t$ and the repulsive on-site interaction between electrons in the
same site with interaction energy $U$.

In the present paper, we purpose to apply this method to an extended Hubbard
model which takes into account the off-site interaction with an interaction
energy $V$

This paper is organized as follows. In sec.2, we present the model and the
calculation procedure which allows us to construct the Hamiltonian matrix of
dimension $85\times 85$. In sec.3, we present our results for ground state
and gap energies and discuss the effect of the off-site interaction on the
dynamics of system. Finally, in Sec.4 we give our conclusions.

\section{\protect\bigskip Model and Formalism}

\bigskip The two-dimensional Extended Hubbard Model on a square lattice is
given by:%
\begin{equation}
H=-\sum_{\left\langle i,j\right\rangle ,\sigma }t_{ij}c_{i,\sigma
}^{^{\dagger }}c_{j,\sigma }+U\sum_{i}n_{i,\uparrow }n_{i,\downarrow
}+V\sum_{\left\langle i,j\right\rangle ,\sigma ,\sigma ^{\prime
}}n_{i,\sigma }n_{j,\sigma ^{\prime }}
\end{equation}

\bigskip where $c_{i,\sigma }^{^{\dagger }}$ ($c_{j,\sigma }$) are the
creation (annihilation) operators for a fermion of spin $\sigma $ at site $i$
( $j$ ) with periodic boundary conditions. Thus, $t$ is the hopping term
from the site $j$ to the site $i$, where $\left\langle i,j\right\rangle $
sums over nearest neighbour sites. The second term describes the local
repulsive interaction with parameter $U$. The last term takes into account
the nearest-neighbour repulsion between electrons with energy $V$.

The resolution of the model ($1$) in case of finite size system, gives the
exact solution of some physical quantities as the ground state energy, the
energy gap and the occupation numbers. We consider, thus, a two dimensional $%
L\times L=N=16$ square lattice at one-eighth filling (four electrons per
cluster), with periodic boundary conditions in both directions. For this
considered system, three types of particle configurations may occur. First,
we can have two double occupancies at sites $i$ and $j$ ($i\neq j$). Second,
we may have a double occupancy at site $i$ and two electrons with opposite
spins at sites $j$ and $k$ ($i\neq j$, $i\neq k$ and $j\neq k$). Finally, we
may have four single occupancies placed on different sites of this $4\times
4 $ square lattice. These three possible configurations provide,
respectively, the three states:

\begin{equation*}
\left\vert a\right\rangle =\left( c_{i,\sigma }^{^{\dagger }}c_{i,-\sigma
}^{^{\dagger }}\right) \left( c_{j,\sigma }^{^{\dagger }}c_{j,-\sigma
}^{^{\dagger }}\right) \left\vert 0\right\rangle
\end{equation*}

\begin{equation*}
\left\vert b\right\rangle =\left( c_{i,\sigma }^{^{\dagger }}c_{i,-\sigma
}^{^{\dagger }}\right) \left( c_{j,\sigma }^{^{\dagger }}c_{k,-\sigma
}^{^{\dagger }}\right) \left\vert 0\right\rangle
\end{equation*}

\begin{equation*}
\left\vert c\right\rangle =\left( c_{i,\sigma }^{^{\dagger }}c_{j,\sigma
}^{^{\dagger }}\right) \left( c_{k,-\sigma }^{^{\dagger }}c_{l,-\sigma
}^{^{\dagger }}\right) \left\vert 0\right\rangle
\end{equation*}

where $\left\vert 0\right\rangle $ represents the state vacuum with no
electron present.

\FRAME{ftbpFU}{6.5276in}{2.0868in}{0pt}{\Qcb{The three possible types of
particle configurations (a: two double occupancies -- b: double occupancy
and two single occupancies -- c: four single occupancies) }}{}{Figure}{%
\special{language "Scientific Word";type "GRAPHIC";maintain-aspect-ratio
TRUE;display "USEDEF";valid_file "T";width 6.5276in;height 2.0868in;depth
0pt;original-width 8.5513in;original-height 2.7086in;cropleft "0";croptop
"1";cropright "1";cropbottom "0";tempfilename
'JQD5NT00.wmf';tempfile-properties "XPR";}}

The states $\left\vert a\right\rangle $ , $\left\vert b\right\rangle $ and $%
\left\vert c\right\rangle $ generate $N_{d}=\frac{16\times 16\times 15\times
15}{2\times 2}=28800$ states, which describe all possible distributions of
our four electrons in the $4\times 4$ cluster. In order to construct a $%
r-space$ basis, it is convient to define a linear operator $T$ \cite{kovacs}
, which verifies the relation:

\begin{equation*}
T(A+B)=T(A)+T(B)
\end{equation*}

\bigskip Where $A$ and $B$ represent the particle configurations as
mentioned in Fig. 1, $T(A)$ is the linear combination of the $16$
contributions obtained by the translation of the configuration $A$ to each
site of the $4\times 4$ cluster. Using this linear operator and taking into
account the symmetry proprieties of the $4\times 4$ cluster in $r-space$
representation, we can regroup these $N_{d}$ states in $85$ cluster states
denoted by $\left\vert n\right\rangle $, where $n=1,2,\ldots ..,85$, and are
all orthogonal vectors. For example, $\left\vert 1\right\rangle $ is the
linear combination of all states $\left\vert a\right\rangle $ with $%
\left\vert R_{i}-R_{j}\right\vert =a$, where $R_{i}$ ($R_{j}$) is the
lattice position of the site $i$ ($j$) and $a$ is the square lattice
parameter. $\left\vert 2\right\rangle $ is the linear combination of all
states $\left\vert b\right\rangle $ with $\left\vert R_{i}-R_{j}\right\vert
=\left\vert R_{j}-R_{k}\right\vert =a$ and $\left\vert
R_{i}-R_{k}\right\vert =2a$. Whereas $\left\vert 3\right\rangle $ is ,also,
the linear combination of all states $\left\vert b\right\rangle $ with $%
\left\vert R_{i}-R_{j}\right\vert =\left\vert R_{j}-R_{k}\right\vert =a$ but 
$\left\vert R_{i}-R_{k}\right\vert =\sqrt{2}a$ \ (For the definition of the
other 82 vectors, see ref. \cite{kovacs}).

The application of the Hamiltonian $(1)$ on the basis vector $\left\vert
1\right\rangle $ gives:

\begin{equation*}
H|1\rangle =\left( 2U+4V\right) |1\rangle +t(|2\rangle -|3\rangle )
\end{equation*}

It is clear that the Hamiltonian would not be diagonal in our $r-space$
basis, since the application of the kinetic term $H_{0}=-\sum
t_{ij}c_{i,\sigma }^{^{\dagger }}c_{j,\sigma }$ on a vector $\left\vert
n\right\rangle $ gives, always, the new states $\left\vert n^{\prime
}\right\rangle $ after the creation and the annihilation of electrons in the
different lattice sites. Thus, it is necessary to define the matrix energy $%
E_{85\times 85}$ as:

\begin{equation*}
E_{nm}=\left\langle n\left\vert H\right\vert m\right\rangle
\end{equation*}

Where $\left\vert n\right\rangle $ and $\left\vert m\right\rangle $ are two
vectors of the $r-space$ basis. In order to study the local properties of
the $4\times 4$ square lattice, it is appropriate to determine ,
numerically, the eigenvalues and the eigenvectors of the matrix energy $%
E_{85\times 85}$. We consider the obtained first and second minimums of the
eigenvalues as, respectively, the ground state and the first excited
energies.

\section{Results and discussion}

First, we disregarded the off-site interaction, and we have plotted in Fig.2
\ and Fig.3 the ground state and the first excited state energies,
respectively, as function of the on-site interaction energy $U$. The
corresponding curves of this case $\left( V=0\right) $ show that the ground
state energy has smooth (less than linear) $U$ dependence, whereas the first
excited state energy is $U$ independent and fixed at $-8t$ for any value of $%
U$. Thus, it is clear that the first excited state is an eigenvector of the
kinetic energy $\left( -\sum_{\left\langle i,j\right\rangle ,\sigma
}t_{ij}c_{i,\sigma }^{^{\dagger }}c_{j,\sigma }\right) $, since this excited
state avoid totally the double occupancy. Whereas, the ground state is not
eigenstate of the kinetic energy or the on-site interacting part of $H$
only, but is eigenvector for the sum of both. Thus, at the ground state, the
dynamics of the electrons system is governed by a competition between the
habitual kinetic and the on-site interaction.

\bigskip Then, we have taken into account the off-site interaction $\left(
V\neq 0\right) $ and we have plotted in in the same previous figures (Fig.2
\ and Fig.3) the ground state and the first excited state energies,
respectively, as function of the on-site interaction energy $U$ for two
values of $V/t$ .

\FRAME{ftbphFU}{4.3967in}{3.1168in}{0pt}{\Qcb{Ground state energy as
function of $U/t$ for different values of $V/t$}}{}{Figure}{\special%
{language "Scientific Word";type "GRAPHIC";maintain-aspect-ratio
TRUE;display "USEDEF";valid_file "T";width 4.3967in;height 3.1168in;depth
0pt;original-width 5.6706in;original-height 4.0136in;cropleft "0";croptop
"1";cropright "1";cropbottom "0";tempfilename
'JQD5NT01.wmf';tempfile-properties "XPR";}}

\FRAME{ftbphFU}{4.1658in}{2.8349in}{0pt}{\Qcb{First excited state energy as
function of $U/t$ for different values of $V/t$}}{}{Figure}{\special%
{language "Scientific Word";type "GRAPHIC";maintain-aspect-ratio
TRUE;display "USEDEF";valid_file "T";width 4.1658in;height 2.8349in;depth
0pt;original-width 5.6074in;original-height 3.8069in;cropleft "0";croptop
"1";cropright "1";cropbottom "0";tempfilename
'JQD5NT02.wmf';tempfile-properties "XPR";}}

The Fig. 2 shows that the energy $E$ has a smooth $U$ dependence and a
linear $V$ dependence. This behaviour is similar to the one obtained for the
ground state energy of the chains Extended Hubbard Model with the Self
Consistent Random Phase Approximation (SCRPA) \cite{ps}, where we have also,
defined the matrix energy but in the basis of vectors impulsion-space ($%
k-space$) and not in $r-space$ representation. The curves of Fig. 2 show,
also, that $E$ decreases with $V$ for a fixed value of $U$. Thus, we can
conclude that this off-site interaction imposes the electron system to avoid
partially the double occupancy in this ground state.

The Fig. 3 shows that the first exited state energy becomes $U$ dependent
for $V\neq 0$. Thus, the off-site interaction encourages the existence of
the double occupancies in this exited state. For a weak off-site interaction
($V/t=0.5$), the corresponding curve shows that $E^{\ast }$ becomes $U$
independent for the high values of $U$. But for an intermediate off-site
interaction ($V/t=1$), $E^{\ast }$ still remains dependent $U$; since, we
have the opportunity to have the double occupancies even for high values of $%
U$.

In order to analyze the on-site and off-site interactions effects on the
repartition of our four electrons in the above system, we define the double
occupancy coefficient in the first exited state $D^{\ast }$ as the
probability to have a couple of electrons $\left( \uparrow \downarrow
\right) $ on the same site in this exited state,

\begin{equation*}
D^{\ast }=\frac{1}{N}\dsum\limits_{i}\left\langle n_{i\uparrow
}n_{i\downarrow }\right\rangle 
\end{equation*}

where the sum over all $4\times 4$ cluster sites and the mean values $%
\left\langle ...\right\rangle $ are taken in the corresponding eigenvector
to $E^{\ast }$.

\FRAME{ftbphFU}{4.2203in}{2.8513in}{0pt}{\Qcb{Double occupancy coefficient $%
D^{\ast }$ in the first exited state as function of $U/t$\ for different
values of $V/t$}}{}{Figure}{\special{language "Scientific Word";type
"GRAPHIC";maintain-aspect-ratio TRUE;display "USEDEF";valid_file "T";width
4.2203in;height 2.8513in;depth 0pt;original-width 5.5374in;original-height
3.7325in;cropleft "0";croptop "1";cropright "1";cropbottom "0";tempfilename
'JQD5NT03.wmf';tempfile-properties "XPR";}}

The double occupancy coefficient $D^{\ast }$ in the first exited state is
shown in Fig.4 as function of $U/t$ for different values of $V/t$. For $%
V/t=0 $, we have $D^{\ast }=0$. Thus, effectively, our system avoids
completely the double occupancy at this exited state. But, after taking into
account the off-site interaction, we have $D^{\ast }\neq 0$. For $V/t=1$, we
have $D^{\ast }\neq 0$ even for the strong values of $U/t$. But, for $%
V/t=0.5 $, the coefficient $D^{\ast }$ vanishes for the strong value of $U/t$%
, where the off-site interaction becomes very weak before the on-site
interaction. Thus, the behaviour of our system in this regime ($U\ll V$) is
similar to the one found in Ref. \cite{kovacs} . Where the authors have
shown that $40\% $ of the excited states of the $4\times 4$ cluster are $U$
independent. But, it is clear that the number of this $U$ independent exited
states decreases if we take into account the off-site interaction, since
this interaction encourage the formation of the double occupancies in the
exited states.

Finally, in order to analyze the effect of the off-site interaction on the
dynamics of electrons, we define the energy gap $\Delta \varepsilon $ as the
difference between the first excited state energy $E^{\ast }$ and the ground
state energy $E$

\begin{equation*}
\Delta \varepsilon =E^{\ast }-E
\end{equation*}

In Fig. 5, we plot the variation of this energy gap $\Delta \varepsilon $ as
function of the repulsive on-site interaction energy $U$ for different
values of the off-site interaction energy $V$.

\FRAME{ftbphFU}{4.2116in}{2.8781in}{0pt}{\Qcb{Energy gap as function of $U/t$
for different values of $V/t$}}{}{Figure}{\special{language "Scientific
Word";type "GRAPHIC";maintain-aspect-ratio TRUE;display "USEDEF";valid_file
"T";width 4.2116in;height 2.8781in;depth 0pt;original-width
5.5711in;original-height 3.7991in;cropleft "0";croptop "1";cropright
"1";cropbottom "0";tempfilename 'JQD5NT04.wmf';tempfile-properties "XPR";}}

For a fixed value of $V/t$, the energy gap decreases with $U$. We deduce
that the repulsive on-site interaction ($U\succ 0$) increases the
conductivity of the system, since the repulsion between the two electrons of
the same site encourages every electron to jump to the neighbouring site.
Moreover, this curves show that the off- site interaction increases also the
conductivity of this $4\times 4$ cluster. For a weak off-site interaction ($%
V/t\preceq 0.5$), the effect of $V$ is remarkable only for the weak on-site
interaction whereas it becomes practically non-existent for strong on-site
interaction. But, for an intermediate off-site interaction ($V/t=1$), the
effect remains remarkable even for the strong values of $U/t$. Thus, we can
conclude that with a strong on-site and an intermediate off-site
interactions, our $4\times 4$ cluster have an important conductivity.

\section{Conclusion}

In this paper, an exact diagonalization in the $r-space$ was proposed to
solve the two dimensional Extended Hubbard Model for finite size system. In
particular, we have considered a two dimensional $L\times L=N=16$ square
lattice at one-eighth filling with respecting the periodic boundary
conditions in both directions.

First, the numeric diagonalization of our Hamiltonian matrix allow us to
determine some interesting local properties of our $4\times 4$ square
lattice as: the ground state energy $E_{GS}$, the first excited state energy 
$E^{\ast }$, the gap energie $\Delta \varepsilon $ and the double occupation
number per site $D^{\ast }$. Then, the analysis of the behaviours of these
obtained local properties as function of $U$ and $V$ allows as to study the
distribution and the dynamics of \ the elctrons system in two interesting
states: ground and first exited states. Thus, we have found that our system
has always the double occupancies in the ground state for any value of $U$
and $V$. In the first exited state, we have shown that the off-site site
interaction encourages the electrons system to form the double occupancies,
where the coefficient $D^{\ast }$ vanishes for $V=0$. Whereas, for an
intermediate off-site interaction, we found that we have always the
probability to have these double occupancies. But, this probability vanishes
for strong on-site and weak off-site interactions, where, the electrons
system avoids completely the double occupancy. Thus, the behaviour of our
system in this regime ($V\ll U$) is similar to the one found in Ref. \cite%
{kovacs}. Finally, the analysis of the off-site interaction effect on the
energy gap shows that the repulsive off-site interaction induces
supplementary conductivity of the system, where, this effect of $V$ is more
remarkable for an intermediate off-site interaction, since we have a
reduction of order $20\%$ in this regime.

\end{document}